\documentclass{article}
\usepackage[T1]{fontenc} 
\usepackage[utf8]{inputenc} 
\usepackage{ismir,amsmath,amssymb,cite,url}
\usepackage{graphicx}
\usepackage{color}
\usepackage{array, multirow, booktabs}
\usepackage{float}
\usepackage{bm}
\usepackage{xurl}

\usepackage{lineno}

\title{Pitch-informed instrument assignment using a deep convolutional network with multiple kernel shapes}

\multauthor
{Carlos Lordelo$^{1,2}$ \hspace{1cm} Emmanouil Benetos$^1$ \hspace{1cm} Simon Dixon$^1$  \hspace{1cm} Sven Ahlbäck$^2$} {
$^1$ Queen Mary University of London, UK \hspace{1cm} $^2$ Doremir Music Research AB, Sweden\\
}

\def\authorname{C. Lordelo, E. Benetos, S. Dixon, and S. Ahlbäck}

\begin{document}

\maketitle

\begin{abstract}
This paper proposes a deep convolutional neural network for performing note-level instrument assignment. Given a polyphonic multi-instrumental music signal along with its ground truth or predicted notes, the objective is to assign an instrumental source for each note. This problem is addressed as a pitch-informed classification task where each note is analysed individually. We also propose to utilise several kernel shapes in the convolutional layers in order to facilitate learning of efficient timbre-discriminative feature maps. Experiments on the MusicNet dataset using $7$ instrument classes show that our approach is able to achieve an average F-score of $0.904$ when the original multi-pitch annotations are used as the pitch information for the system, and that it also excels if the note information is provided using third-party multi-pitch estimation algorithms. We also include ablation studies investigating the effects of the use of multiple kernel shapes and comparing different input representations for the audio and the note-related information.
\end{abstract}
\section{Introduction}\label{sec:intr}
Automatic Music Transcription (AMT) is the process of creating any form of notation for a music signal and is currently one of the most challenging and discussed topics in the Music Information Retrieval (MIR) community \cite{Benetos2019}. 
Most AMT systems are designed to transcribe a single monophonic or a single polyphonic source into a musical score (or piano-roll). In this case, the main sub-task involved in the process is Multi-Pitch Estimation (MPE), where predictions regarding the pitch and time localisation of the musical notes are carried out. However, when analysing polyphonic multi-instrumental recordings, not only each note should have its pitch and duration properly estimated, but the information regarding the timbre of sounds should also be correctly processed \cite{Duan2014}. It is mandatory to have a way of recognising the instrument that played each note. 

In this paper, we propose a pitch-informed instrument assignment approach, where the main objective is to associate each note event of a music signal to one instrument class. In contrast to other state-of-the-art instrument recognition approaches, which are usually addressed on a frame-level \cite{Hung2018, Hung2019} or clip-level \cite{Han17, Gururani2019} basis, our approach analyses each note event individually. Therefore, it is possible to say that we perform a note-level instrument recognition.

Previous work has shown that the use of pitch information can help frame-level instrument recognition \cite{Hung2018}. Inspired by this, we propose a framework that uses an auxiliary input based on note-event pitch information. Our system is trained using the note annotations provided in the MusicNet \cite{Thickstun17} dataset. However, our main motivation is to create a modular framework that can be combined with any MPE algorithm in order to obtain multi-instrumental pitch predictions, which allows for transcribing music in staff notation, corresponding to the perception of pitch events. Therefore, we also show that our approach can obtain good performance when the note information is predicted by state-of-the-art MPE algorithms such as \cite{Thome17, Wu2019a}.



Furthermore, the utilisation of multiple kernel shapes in the filters of a Convolutional Neural Network (CNN) has been proven to be an efficient strategy of applying domain knowledge in several MIR tasks \cite{Pons2016, Pons17_Timbre, Lordelo19}. In particular, \cite{Lordelo19} applied this strategy with a dense connectivity pattern of skip-connections in order to learn even more efficient feature-maps and reduce the number of trainable parameters for the task of source separation. In our work, we build our CNN adapting the architecture in \cite{Lordelo19} for the classification (instrument assignment) task and verified that it can also improve its performance. In summary, the main contributions of this paper are as follows:
\begin{itemize}
    \item Pitch-informed instrument assignment: Proposal of a Deep Neural Network (DNN) that associates each note from a music signal to its instrumental source.
    \item Modular Framework: Approach works with any MPE method. We evaluate the performance when using ground-truth note labels as well as $2$ state-of-the-art MPE algorithms \cite{Thome17, Wu2019a}.  
    \item Multiple Kernel Shapes: Proposal of a CNN architecture for instrument assignment that uses multiple kernel shapes for the convolutions, facilitating learning representations for different instruments and note sound states. We show that their use improves instrument assignment performance.
\end{itemize}


        


\begin{table*}[htb]
\centering
\resizebox{0.978\textwidth}{!}{%
\begin{tabular}{ccccccccccccc}
\toprule
Set & Piano & Violin & Viola & Cello & Horn & Bassoon & Clarinet & Harps. & Bass & Oboe & Flute & Total\\ \toprule
\multicolumn{1}{c}{\multirow{2}*{Train}} 
& $628549$ & $197229$ & $88446$ & $89356$ & $10770$ & $13874$ & $22873$ & $4914$ & $3006$ & $8624$ & $8310$ & $1075951$ \\ \cmidrule{2-13}
& $58.4$\% & $18.3$\% & $8.2$\% & $8.3$\% & $1.0$\% & $1.2$\% & $2.1$\% & $0.5$\% & $0.3$\% & $0.8$\% & $0.8$\% & $100$\% \\ \bottomrule
\toprule
\multicolumn{1}{c}{\multirow{2}*{Test}} 
& $5049$ & $3238$ & $842$ & $1753$ & $557$ & $873$ & $1277$ & $0$ & $0$ & $0$ & $0$ & $13589$ \\ \cmidrule{2-13}
& $37.2$\% & $23.8$\% & $6.2$\% & $12.9$\% & $4.1$\% & $6.4$\% & $9.4$\% & $0$ & $0$ & $0$ & $0$ & $100$\% \\ \bottomrule
\end{tabular}%
}
\caption{Statistics of the note labels provided by MusicNet across train and test sets.}
\label{tab:musicnet}
\end{table*}

\section{Related Work}
The instrument recognition task is usually formulated as a multi-label classification task that can be addressed either on a frame-level \cite{Hung2018, Hung2019, Wu2020}, where the purpose is to obtain the instrument activations across time, or on a clip-level basis \cite{Han17, Gururani2019, Solanki2019}, where the purpose is to estimate the instruments that are present in an audio clip. However, our objective in this work is to approach the instrument recognition task note by note, assigning an instrument class to each. Such a task requires note-event annotations and, in the literature, it is also known as \emph{instrument assignment} \cite{Benetos2013} or \emph{multi-pitch streaming} \cite{Duan2014}. 

Just few works have explored this particular task. For instance, Duan et al.\ \cite{Duan2014} approached it using a constrained clustering of frame-level pitch estimates obtained from an MPE algorithm via the minimisation of timbre inconsistency within each cluster. They tested different timbre features for both music and speech signals. In \cite{Arora2015}, a similar method was proposed, where the authors applied Probabilistic Latent Component Analysis (PLCA) to decompose the audio signal into multi-pitch estimates and to extract source-specific features. Then, clustering was performed under the constraint of cognitive grouping of continuous pitch contours and segregation of simultaneous pitches into different source streams using Hidden Markov Random Fields. Both of those works, however, assume that each source is monophonic, i.e., each instrument could only play a single note at a time. 

An alternative approach iss to model the temporal evolution of musical tones \cite{Benetos2013}. This method is based around the use of multiple spectral templates per pitch and instrument source that correspond to sound states. The authors used hidden Markov model-based temporal constraints to control the order of the templates and streamed the pitches via shift-invariant PLCA. In a more recent work, Tanaka et al.\ \cite{Tanaka2020} also approached the task via clustering, but applied on a joint input representation combined of the spectrogram and the pitchgram, which was obtained using an MPE algorithm. In their proposal, each bin of the joint input was encoded onto a spherical latent space taking into account timbral characteristics and the piano-rolls of each instrument were later estimated via masking of the pitchgram based on the results of a deep spherical clustering technique applied on the latent space.

Recent multitask deep-learning based works have successfully proposed multi-instrumental AMT methods that are able to directly estimate pitches and associate them to their instrumental source jointly \cite{Bittner2018, Hung2019, Manilow2020}. In \cite{Bittner2018}, a multitask deep learning network jointly estimated outputs for various tasks including multiple-pitch, melody, vocal and bass line estimation. The Harmonic Constant-Q Transform (HCQT) of the audio signal was used as input and the data used for training was semi-automatically labelled by remixing a diverse set of multitrack audio data from the MedleyDB \cite{Bittner2017} dataset. In \cite{Hung2019} a DNN was used to jointly predict the pitch and instrument for each audio frame. They used the Constant-Q Transform (CQT) as input to their system and trained using a large amount of audio signals synthesised from MIDI piano-rolls. Manilow et al.\  \cite{Manilow2020}, on the other hand, were able to jointly transcribe and separate an audio signal into up to $4$ instrumental sources --- piano, guitar, bass and strings. 
However, their system was trained with only synthesised signals.

Our approach is closely related to that of Hung and Yang \cite{Hung2018}, where a frame-level instrument recogniser is proposed using the CQT spectrogram of the music signal allied with the pitch information of the note events. We also use the pitch annotations to guide the instrument classifier, but our work differs from \cite{Hung2018} in the fact that we perform a classification for each note event individually, while Hung and Yang use the whole piano-roll at once to guide frame-level instrument recognition. While Hung and Yang are able to obtain the instrument activations leveraging from the pitch information, they cannot stream the note events into their corresponding instruments.

\section{Proposed Method}\label{sec:method}
In our method, we use the same definition of note events as in the MIREX MPE task\footnote{\url{https://www.music-ir.org/mirex/}}. Each note $N$ is considered an event with a constant pitch $f_0$, an onset time $T_\mathrm{on}$ and an offset time $T_\mathrm{off}$. Therefore, if a music signal has a total of $M$ notes, any note $N_i$, with $i \in \{1, \cdots, M\}$, can be uniquely defined by the tuple $(f_{0}^{i}, T_\mathrm{on}^i, T_\mathrm{off}^i)$. In our experiments, we use two ways of obtaining this note information. The first using ground-truth pitch labels provided by the employed dataset (MusicNet) \cite{Thickstun17} and the second using pitch estimates predicted by state-of-the-art MPE algorithms \cite{Thome17, Wu2019a}. We consider the $f_0$ granularity to follow the semitone scale, ranging from $A0$ to $G\sharp7$ (MIDI \#$21-104$).

\begin{figure}[t]
    \centerline{
    \includegraphics[width=0.85\columnwidth]{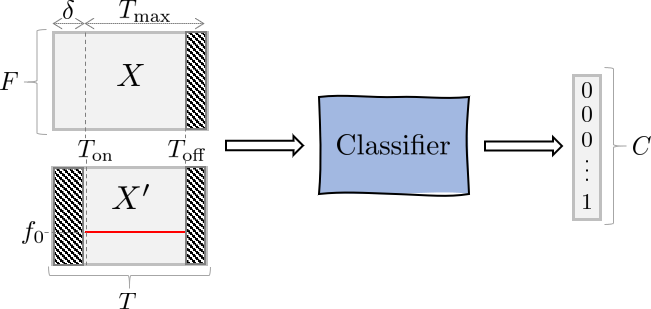}}
    \caption{Overview of the proposed framework for instrument assignment. See Section \ref{sec:method} for the detailed explanation of the variables in the figure.}
    \label{fig:inst_track}
\end{figure}

In our framework polyphony is allowed, so, most of the time more than a single note will be active, but our objective is to analyse each note of the audio signal separately in order to be able to assign an instrument class to it. This is done by using two inputs to the model: the main input $X(f,t)$, with $f$ representing frequency and $t$ representing time, is a time-frequency representation of a segment of the audio signal around the value of $T_\mathrm{on}$, and an auxiliary input $X'(f,t)$, which carries information regarding $f_0$, $T_\mathrm{on}$ and $T_\mathrm{off}$. 
The two inputs are concatenated into a two-channel input $\mathbf{X}(f,t,c)$, where $c\in\{1,2\}$ represents the channel dimension, that is fed to the model. In Figure \ref{fig:inst_track} an overview of the proposed framework is shown.

\subsection{Main Audio Input}
The main input is a time-frequency representation $X(f,t) \in \mathbb{R}^{F \times T}$ of a small clip of the music signal, where $F$ is the number of frequency bins and $T$ is the number of time frames. The clip is generated by first setting a maximum duration $T_\mathrm{max}$ for the note. 
We tested values of $T_\mathrm{max}$ ranging from $400$\,ms to $1$\,s (see Section \ref{sec:results} for details) and $400$\,ms obtained the best results, so we kept this value in all of the other experiments. If any note $N_i$ has a duration $D_i$ greater than $T_\mathrm{max}$, i.e., $D_i = T_\mathrm{off} - T_\mathrm{on} > T_\mathrm{max}$, only its initial time span of $T_\mathrm{max}$ seconds is considered.

Next, for every note $N_i$, $X_i(f,t)$ is constructed by picking a segment of duration $T = T_{\mathrm{max}} + \delta$ from the original music signal starting from $T_{\mathrm{on}}^{i} - \delta$, where $\delta$ is a small interval to take into account deviations between the true onset value and the value we use. The inclusion of the extra window of $\delta$ from the music signal also helps the convolutional layers since it brings some context of the signal before the note onset value. We set $\delta = 30$~ms after initial tests. Lastly, if the note duration $D_i$ is less than $T_{\mathrm{max}}$, we set the values of $X(f, t > D_i)$ to zero, where $D_i = T_{\mathrm{off}}^{i} - T_{\mathrm{on}}^{i}$.

\subsection{Auxiliary Note-Related Input}
\begin{figure}[t]
    \centerline{
    \includegraphics[width=0.85\columnwidth]{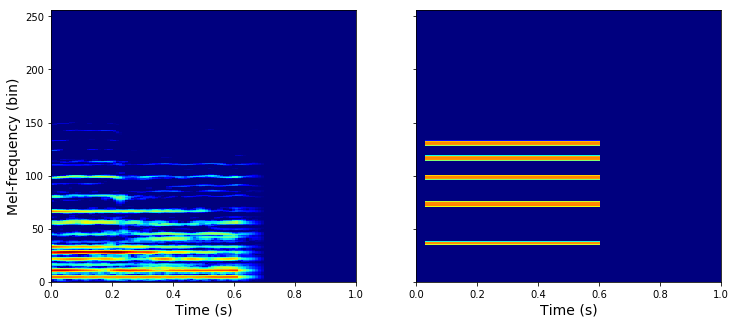}}
    \caption{A pair of inputs using $256$ mel-frequency spectrogram. In the left is depicted $X(f,t)$, where three pitches are simultaneously activated (MIDI \# $58$, $62$ and $74$) and in the right $X'(f,t)$, where the note with pitch \# $74$, is modelled using an harmonic comb of $H=5$. 
    In this example, $D_i\!=\!600$\,ms and $T_\mathrm{max}\!=\!1$\,s.}
    \label{fig:inputs}
\end{figure}

The auxiliary input $X'(f,t) \in \mathbb{R}^{F \times T}$ is a harmonic comb representation using the pitch value $f_0$ as the first harmonic\footnote{We use the definition that $f_0$ corresponds to the first harmonic.}, such that,     
\begin{equation}
    X'(f,t) = \begin{cases}
      1, \quad \mathrm{if} ~\; f= hf_0 ~\; \mathrm{and}~\; T_\mathrm{on} \leq t \leq T_\mathrm{off}\\
      0, \quad \mathrm{otherwise}\\
    \end{cases},
\end{equation}
where $h=\{1,2,3,\cdots, H\}$ with $H$ being the total number of harmonics in the representation. We tested multiple values for $H$ (see Section \ref{sec:exp}). In practice, we use a tolerance of half a semitone for each harmonic value when constructing $X'(f,t)$ as a mel-spectrogram. Therefore, even though this representation starts as binary, the final mel-spectrogram is not binary due to the mel-filtering procedure. Moreover, it is important to note that we also set the values of $X'$ before $T_\mathrm{on}$ and after $T_\mathrm{off}$ to zero. In Figure \ref{fig:inputs} we show an example of a pair of inputs for our framework.

\subsection{Output}
The note-level instrument assignment task is tackled as a multi-class single-label classification task. Given $\mathbf{X}$, our objective is to classify it as belonging to one of $C$ instrument classes. We use a deep neural network that receives $\mathbf{X}$ as input and outputs a $C$-dimensional vector $\hat{y}$. A softmax activation function is applied in the final layer of the network to ensure the values of $\hat{y}$ represent probabilities that sum up to $1$. The model is trained using the cross-entropy loss. At inference time, the class corresponding to the dimension with the highest value in $\hat{y}$ is predicted. See Section \ref{sec:arch} for details regarding the network architecture.


In the cases where two or more instruments are playing the same pitch simultaneously, the small differences between the notes' onset and offset values can generate different inputs $\mathbf{X}$. Thus, it would still allow the instrument assignment task to be properly executed as a single label classification scenario. However, when the pitch, onset and offset values of notes from different instruments exactly match, our system will consider them as a single note and only a single instrument will be estimated. This case rarely happens in real-world scenarios for many musical styles. For instance, in MusicNet
only $0.9\%$ of the notes had the same pitch, onset, and offset values.
For our experiments, we have considered notes in MusicNet that were performed by a single instrument, and discarded the notes that were concurrently produced (in terms of the same pitch, onset, and offset times) by multiple instruments. As a proof of concept, we believe that this is not a severe limitation for our framework and we leave multi-labelled approaches as future work.



\begin{figure*}[htb]
    \centerline{
    \includegraphics[width=0.90\textwidth]{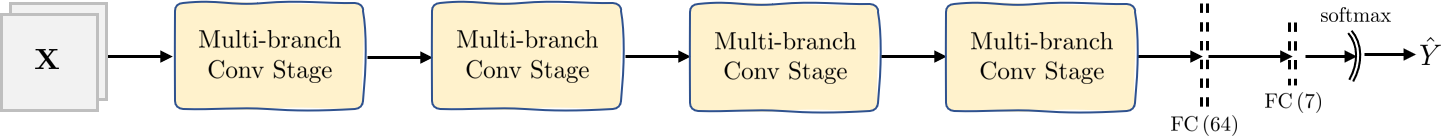}}
    \caption{Proposed network architecture. FC represents a Fully Connected layer with LeakyRelu activation function.}
    \label{fig:arch_full}
\end{figure*}


\begin{table*}[htb]
\centering
\resizebox{0.9\textwidth}{!}{%
\begin{tabular}{cccccccccc}
\toprule
Main Input & Aux Input & Piano & Violin & Viola & Cello & Horn & Bassoon & Clarinet & Mean\\ \toprule
\multicolumn{1}{c}{\multirow{8}*{CQT}} 
& --- & $0.960$ & $0.732$ & $0.116$ & $0.725$ & $0.512$ & $0.296$ & $0.681$ & $0.575$\\[-0.4pt] \cmidrule{2-9}
& $H=1$ & \bm{$0.994$} & $0.934$ & $0.763$ & $0.955$ & $0.783$ & \bm{$0.888$} & $0.950$ & \bm{$0.895$}\\[-0.4pt] \cmidrule{2-9}
& $H=2$ & $0.993$ & $0.939$ & $0.771$ & \bm{$0.960$} & \bm{$0.785$} & $0.858$ & $0.929$ & $0.891$\\[-0.3pt] \cmidrule{2-9}
& $H=3$ & $0.992$ & \bm{$0.946$} & \bm{$0.772$} & $0.957$ & $0.754$ & $0.884$ & \bm{$0.952$} & $0.894$\\[-0.4pt] \cmidrule{2-9}
& $H=4$ & $0.993$ & $0.938$ & $0.766$ & $0.958$ & $0.784$ & $0.869$ & $0.950$ & $0.894$\\[-0.4pt] \cmidrule{2-9}
& $H=5$ & $0.993$ & $0.939$ & $0.767$ & $0.952$ & $0.769$ & $0.874$ & $0.949$ & $0.892$\\[-0.4pt] \midrule
\multicolumn{1}{c}{\multirow{8}*{\begin{tabular}{c}
Mel\\
STFT
\end{tabular}
 }}
& --- & $0.967$ & $0.742$ & $0.222$ & $0.730$ & $0.607$ & $0.306$ & $0.690$ & $0.609$\\[-0.4pt] \cmidrule{2-9}
& $H = 1$ & $0.996$ & $0.939$ & $0.759$ & $0.958$ & $0.780$ & $0.867$ & $0.958$ & $0.895$\\[-0.4pt]\cmidrule{2-9}
& $H = 2$ & $0.994$ & $0.945$ & $0.779$ & $0.956$ & $0.809$ & $0.864$ & $0.946$ & $0.899$\\[-0.4pt]\cmidrule{2-9}
& $H = 3$ & \bm{$0.997$} & $0.944$ & $0.775$ & \bm{$0.958$} & $0.8104$ & $0.879$ & \bm{$0.967$} & \bm{$0.904$}\\[-0.4pt] \cmidrule{2-9}
& $H = 4$ & $0.996$ & $0.935$ & $0.747$ & $0.945$ & \bm{$0.839$} & \bm{$0.891$} & $0.960$ & $0.902$\\[-0.4pt] \cmidrule{2-9}
& $H = 5$ & $0.996$ & \bm{$0.947$} & \bm{$0.783$} & $0.954$ & $0.801$ & $0.876$ & $0.954$ & $0.902$\\[-0.4pt] \bottomrule
\end{tabular}%
}
\caption{Evaluation of instrument assignment task when using CQT or Mel spectrograms as input representation for the network as well as a comparison between models trained with no auxiliary input and models trained with different number of harmonics in the auxiliary input. This experiment was performed using $T_{max} = 400$\,ms}
\label{tab:results_input}
\end{table*}

\section{Architecture}\label{sec:arch}
When processing music spectrograms by CNNs, the strategy of combining vertical and horizontal kernel shapes in the model architecture can facilitate learning of timbre-discriminative feature maps \cite{Pons2016, Pons17_Timbre, Lordelo19}. In our work, we propose a CNN adapted from the $3$W-MDenseNet \cite{Lordelo19}. This architecture was originally proposed for harmonic-percussive source separation and consists of an encoder-decoder model that estimates spectrograms for two sources. Thus, the outputs of the $3$W-MDenseNet have the same shape as the mixture spectrogram that is used as input. In this CNN architecture, three MDenseNets \cite{MDenseNet17} run in parallel in separated branches, each with a unique kernel shape (vertical, square and horizontal). The MDenseNets are only combined at the final layer, i.e., after both the encoding and decoding procedure are performed. In our work, we adopt a similar methodology by taking only the encoder layers from \cite{Lordelo19} and adding fully connected layers at the end in order to perform classification rather than separation. Also, we propose modifications to the original encoder layers: instead of combining the branches using a concatenation layer only at the final stage, we concatenate their feature-maps at the end of each downsampling stage. By doing so, we allow each branch to have access to feature-maps computed using all different choices of kernel shapes from a previous stage. 

\begin{figure}[htb]
    \centerline{
    \includegraphics[width=0.99\columnwidth]{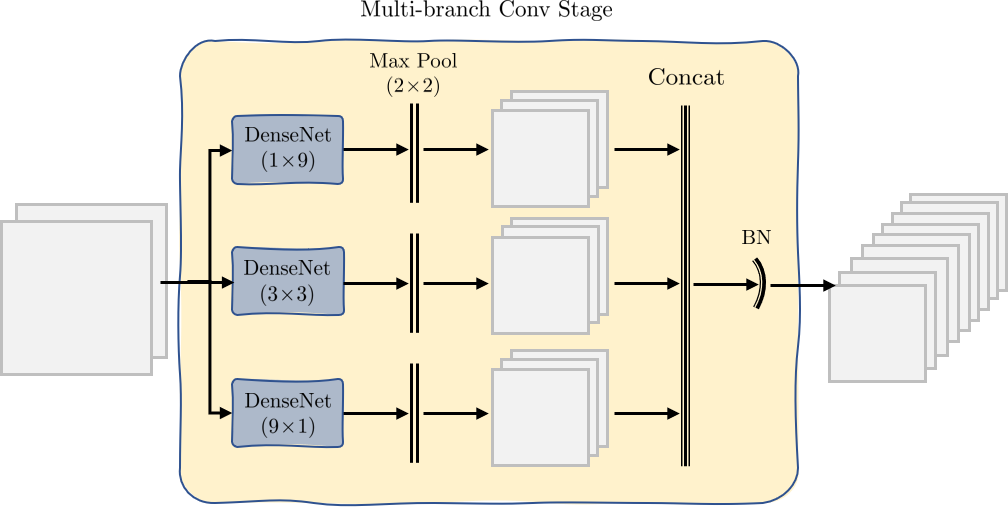}}
    \caption{Internal structure of a multi-branch conv. stage. Each DenseNet has a growth rate $k\!=\!25$ and $L\!=\!4$ layers. BN represents a Batch Normalisation layer. We use LeakyRelu as activation function after each conv. layer.}
    \label{fig:arch_detail}
\end{figure}

Figure \ref{fig:arch_full} shows a summary of the architecture we adopt in our work. It consists of a stack of $4$ multi-branch convolutional stages and $2$ fully connected layers. In Figure \ref{fig:arch_detail} the internal structure of the multi-branch convolutional stage is shown. Internally, each multi-branch convolutional stage contains $3$ separate branches whose convolutions have unique kernel shapes. We use a branch with horizontal ($1\!\times\!9$), a branch with square ($3\!\times\!3$), and a branch with vertical ($9\!\times\!1$) convolutions. In each path, a Densely connected convolutional Network (DenseNet) \cite{Huang2017_DenseNet} with growth rate $k\!=\!25$ and number of layers $L\!=\!4$ is used. In short, a DenseNet is a stack of $L$ $k$-channel convolutional layers --- each with its own activation function --- with a dense pattern of skip connections, where each layer receives the concatenation of all previous layers' outputs as input. We used the LeakyRelu function as the activation function for all layers. The reader is referred to \cite{Huang2017_DenseNet} for the detailed internal structure of a DenseNet. After the DenseNet, a $(2\times2)$ max pooling layer is applied in order to reduce the feature-maps' dimensions and increase the receptive field at each branch. Afterwards, the three branches are concatenated and the batch is normalised. The final feature-maps are used as input for the next multi-branch convolutional stage. Since we need to concatenate feature-maps that were originated by multiple kernel shapes we use padding on the convolution and on the max pooling to ensure the feature-maps maintain the same dimensions across branches. The number of trainable parameters is approximately $1.1$ million.

\section{Dataset}\label{sec:data}

We used the MusicNet dataset \cite{Thickstun17} in our experiments. MusicNet is the largest publicly available dataset with non-synthesised data that is strongly labelled for the task of instrument recognition. This means that we know the exact frames where the instruments are active in the signal, which permits the training of supervised models to perform instrument recognition at the frame-level, note-level, and clip-level. The dataset contains $330$ freely-licensed classical music recordings by $10$ composers, written for $11$ instruments, along with over $1$ million annotated labels indicating the precise time and pitch of each note in the recordings and the instrument that plays each note. 

The instrument taxonomy of MusicNet is: piano, violin, viola, cello, french horn, bassoon, clarinet, harpsichord, bass, oboe and flute. However, the last $4$ instruments (harpsichord, bass, oboe and flute) do not appear in the original test set provided by the authors. Therefore, in all our experiments we ignored all the labels related to those instruments and we performed a $7$-class instrument classification using the following classes: piano, violin, viola, cello, french horn, bassoon and clarinet. Table \ref{tab:musicnet} shows the statistics of the note labels provided by MusicNet. The dataset is heavily biased towards piano and violin given their usual presence in Western classical music recordings. 




\section{Experimental Setup}\label{sec:exp}
In all experiments we used the original train/test split provided by MusicNet with the original sampling frequency of $44100$~Hz. For experiments that involved the computation of Short Time Fourier Transform (STFT) we used Blackman-Harris windows of $4096$ samples to compute the Discrete Fourier Transform (DFT). The hop size was always set to $10$\,ms in every experiment.

From the training set we picked $5$\% of the notes of each class and created a validation set. We trained the models using the Adam optimiser with an initial learning rate of $0.001$ and reduced it by a factor of $0.2$ if the cross-entropy loss stopped improving for $2$ consecutive epochs on the validation set. If no improvement happened after $10$ epochs, the training was stopped early. The experiments were performed using the Tensorflow/Keras Python package.

The classification performance was evaluated by computing the note-level F-score ($F_\mathrm{s}$), which is directly related to the precision ($P$), recall ($R$) according to: \begin{equation}
    P = \frac{\mathrm{TP}}{\mathrm{TP} + \mathrm{FP}}, \hfill~ R = \frac{\mathrm{TP}}{\mathrm{TP} + \mathrm{FN}}, \hfill~ F_\mathrm{s} = \frac{2PR}{P+R}
\end{equation} where $\mathrm{TP}$ is the number of true positives, $\mathrm{FP}$ the false positives and $\mathrm{FN}$ the false negatives.

For the cases when the instrument assignment is done on top of MPE algorithms, we provide $2$ groups of metrics that are generated following the MIREX evaluation protocol for the music transcription task. In the first group, an estimated note is assumed correct if its onset time is within $50$\,ms of a reference note and its pitch is within quarter tone of the corresponding reference note. The offset values are ignored. In the second group, on top of those requirements, the offsets are also taken into consideration. An estimate note is only considered correct if it also has an offset value within $50$\,ms or within $20$\% of the reference note's duration around the original note's offset, wherever is largest. After all notes are verified, the F-score is computed note-wise across time and the average value is provided here. This evaluation method was computed using the \texttt{mir\_eval.transcription}\footnote{\url{https://craffel.github.io/mir_eval/}} toolbox.

\section{Results}\label{sec:results}
\subsection{Effects of the Kernel Shapes}\label{subsec:result_kernel}
First we analysed the effects of the inclusion of multiple kernel shapes in the architecture of the CNN. The top part of Table \ref{tab:results_kernel_input} compares $3$ versions of the model: one that uses only square filters in a single branch; a version using the branched structure, but with $(3\times3)$ kernels in each; and another model with the proposed multi-branch structure with horizontal, square, and vertical kernels. For the single-branched case we increased the growth factor of the DenseNets to $57$ channels in order to keep the number of trainable parameters of the network close to the original.

Analysing the results we see that the addition of new kernel shapes improved the average F-score across all classes. Regarding each instrument class, we can say that for string instruments (piano, violin, viola and cello) there is a gain in performance, while for non-string instruments (horn, bassoon, clarinet) the performance either drops or remains with a negligible gain if compared to the models that used only square filters. This suggests that the inclusion of vertical kernel shapes helped the model in learning the percussive characteristics of the timbre of string musical instruments.
\begin{table}[htb]
\centering
\resizebox{0.99\columnwidth}{!}{%
\begin{tabular}{ccccccccc}
\toprule
Kernel / $T_{\mathrm{max}}$ & Piano & Violin & Viola & Cello & Horn & Bassoon & Clarinet & Mean\\ \toprule
$(3\!\times\!3)$ & $0.994$ & $0.936$ & $0.757$ & $0.954$ & \bm{$0.826$} & $0.864$ & $0.954$ & $0.898$\\ \midrule
 $3\!\times\!(3\!\times\!3)$& $0.995$ & $0.939$ & $0.764$ & $0.945$ & $0.819$ & \bm{$0.896$} & $0.965$ & $0.903$\\ \midrule
Multiple & \bm{$0.997$} & \bm{$0.944$} & \bm{$0.775$} & \bm{$0.958$} & $0.810$ & $0.879$ & \bm{$0.967$} & \bm{$0.904$} \\
\midrule[1.5pt]
$400$~ms & \bm{$0.997$} & \bm{$0.944$} & \bm{$0.775$} & \bm{$0.958$} & $0.810$ & $0.879$ & \bm{$0.967$} & \bm{$0.904$}\\ \midrule
$600$~ms & $0.996$ & $0.942$ & $0.771$ & $0.954$ & \bm{$0.826$} & \bm{$0.881$} & $0.959$ & $0.904$\\\midrule
$800$~ms & $0.996$ & $0.944$ & $0.772$ & $0.957$ & $0.814$ & $0.868$ & $0.965$ & $0.902$\\ \midrule
$1$~s & $0.997$ & $0.931$ & $0.740$ & $0.954$ & $0.742$ & $0.871$ & $0.948$ & $0.883$\\ \bottomrule[1.0pt]
\end{tabular}%
}
\caption{Instrument assignment performance based on the kernel shapes used in network (first three rows) and based on the value used for the maximum valid note duration $T_{\mathrm{max}}$. The metric shown is the F-score achieved by each class and the average value across all instruments.}
\label{tab:results_kernel_input}
\end{table}

\subsection{Evaluation of the Input Size}
We also tested different values for the input size. More specifically, we compared multiple values for $T_{\mathrm{max}}$, which is the maximum valid window of analysis for a note event. The results are shown in the lower part of Table \ref{tab:results_kernel_input}. We can see that the shortest input size of $400$~ms obtained the best results. We believe that it is due to the fact that the average duration of a note event in the test set of MusicNet is $260$~ms and the $90$th percentile is $0.464$\,ms. So, the value of $400$\,ms is already enough to represent the vast majority of the notes. Moreover, when the analysed note event is longer than $400$\,ms, the $400$\,ms initial window contains most of the important features for the model.


\subsection{Auxiliary Input and Types of Representations}
To test the importance of the auxiliary input and how its modification would affect the performance of the model, we also tested a version of the model using only the main mel spectrogram input and versions using different numbers of harmonics $H$ in the auxiliary input (from $H=1$ to $H=5$). We also tested two types of input representation for the model, the Constant-Q Transform (CQT) and the mel-frequency spectrogram. The CQT was computed using $12$ bins per octave and a total of $115$ bins starting from $G\sharp0$ (MIDI $\#20$). The mel-frequency spectrogram was computed by a linear transformation of an STFT onto a mel-scaled frequency axis, using $256$ mel-bins. The results are provided in Table \ref{tab:results_input}. 

Analysing the results, it is possible to say that the auxiliary input is extremely necessary for the framework. Without it, the average F-score only reaches $60.9$\%, while with it the performance improves up to $90.4$\%. Apart from piano, all other classes have a large decrease in performance when we exclude the auxiliary input from $\mathbf{X}$. We believe that the results for the piano class continue to be high not only because of the MusicNet bias towards piano, but also because some recordings of the test set are solo piano recordings, which facilitates the classification of piano notes when analysing the main input signal due to the absence of other classes. Regarding the number of harmonics used in the auxiliary input, we can see that, in general, the CQT works best with few harmonics, while the Mel-STFT prefers higher values. A possible explanation for this is the fact that it is harder to represent odd harmonics on the CQT using a log-frequency resolution of $12$ bins per octave. However, more experiments are needed in order to better investigate this assumption. 

\subsection{Streaming of Multi-Pitch Estimations}
Once we verified that our model obtains impressive performance when the original ground-truth labels are used, we tested the classifier in a more realistic environment, where no note-event labels are readily available. We estimated frame-level pitch values using two third-party MPE algorithms \cite{Thome17, Wu2019a}. For the algorithm in \cite{Thome17} we obtained an implementation from the original authors while an implementation of \cite{Wu2019a} is available via the project Omnizart\footnote{\url{https://github.com/Music-and-Culture-Technology-Lab/omnizart}}. We ran both algorithms on the music recordings to obtain the note events in order to construct the input to the classifier.

It is important to observe that errors in the MPE estimation will be carried over to the instrument assignment task. If a note is wrongly estimated, no ground-truth class for the instrument assignment exists, so it is hard to evaluate the results in the same way we did for the other experiments. So, in this experiment we used the transcription metrics that we explained in the last paragraph of Section \ref{sec:exp}. The results appear in Table \ref{tab:mpe}. 
Given the limitations of each MPE method we used, we can see that our approach can successfully generate multi-instrument transcriptions.

\begin{table}[htb]
\centering
\resizebox{0.85\columnwidth}{!}{%
\begin{tabular}{ccccccc}
\toprule
 \multirow{2}{*}{Instr.} & \multicolumn{3}{c}{Onset} & \multicolumn{3}{c}{Onset + Offset} \\[-0.3pt]\cmidrule(lr){2-4} \cmidrule(lr){5-7}
 & GT & \cite{Thome17}  & \cite{Wu2019a} & GT & \cite{Thome17} & \cite{Wu2019a}\\ \toprule
MPE-only & $1$ & $0.633$ &  $0.480$ & $1$ & $0.423$ & $0.200$ \\[-0.3pt] \midrule
piano & $0.942$ & $0.745$ &  $0.451$ & $0.942$ & $0.497$ & $0.196$\\[-0.3pt] \midrule
violin & $0.997$ & $0.529$ & $0.499$ & $0.997$ & $0.381$ & $0.225$\\[-0.3pt] \midrule
viola & $0.775$ & $0.366$ & $0.308$  & $0.775$ & $0.227$ & $0.116$\\[-0.3pt] \midrule
cello & $0.954$ & $0.596$ & $0.570$  & $0.954$ & $0.507$ & $0.258$\\[-0.3pt] \midrule
horn& $0.804$ & $0.460$ & $0.429$ & $0.804$ & $0.232$ & $0.166$ \\[-0.3pt] \midrule
bass. & $0.874$& $0.473$ & $0.373$ & $0.874$ & $0.193$ & $0.130$\\[-0.3pt] \midrule
clar. & $0.967$ & $0.616$ & $0.456$ & $0.967$ & $0.344$ & $0.165$\\[-0.3pt] \midrule
\end{tabular}%
}
\caption{Transcription results when using Ground-Truth (GT) labels and when using two different MPE methods. In the row "MPE-only" no instrument assignment is done, we evaluate the multi-pitch estimates using the reference ground-truth notes ignoring the instrument annotations.}
\label{tab:mpe}
\end{table}

\section{Conclusions}
In this work we presented a convolutional neural network for note-level instrument assignment. We approach this problem as a classification task and proposed a framework that uses the pitch information of the note-events to guide the classification. Our approach can also successfully classify notes provided by a MPE algorithm, which permits generating multi-instrument transcriptions. Our method also shows the benefits of including different kernel shapes in the convolutional layers.

As future work we plan to investigate more deeply the interaction of our method with MPE algorithms as well as how the final estimations can be improved by including a clip-level analysis. The adoption of multi-label classification approaches is also planned. 

\section{Acknowledgement}
This work is supported by the European Union's Horizon 2020 research and innovation programme under the Marie Skłodowska-Curie grant agreement No.~765068 (MIP-Frontiers).

\bibliography{ref}

\end{document}